\documentclass[journal]{IEEEtran}
\usepackage[left=0.75in, right=0.75in, top=1in, bottom=0.9in]{geometry}
\usepackage{mathrsfs,amsthm}
\usepackage{cite}
\usepackage{amsmath,amssymb,mathrsfs,amsthm,bm}

\makeatletter
\def\ps@headings{%
\def\@oddhead{\mbox{}\scriptsize\rightmark \hfil \thepage}%
\def\@evenhead{\scriptsize\thepage \hfil \leftmark\mbox{}}%
\def\@oddfoot{}%
\def\@evenfoot{}}
\makeatother 
\usepackage{amsmath,amssymb,stfloats,subfigure,multicol}
\usepackage{float}
\usepackage{graphicx}
\usepackage{epsfig,epsf,psfrag,amssymb,amsfonts,latexsym,graphicx,mathrsfs,subfigure}
\usepackage{bbm}
\usepackage{dsfont}
\usepackage{chngpage}
\usepackage[dvips]{color}
\usepackage{textcomp}
\usepackage{hyperref}
\usepackage{url}
\usepackage[hyphenbreaks]{breakurl}
\usepackage[normalem]{ulem}

\usepackage{times}
\usepackage{fancyhdr,graphicx,amsmath,amssymb}
\usepackage{amsmath}

\newtheorem{theorem}{Theorem}

\newtheorem{lemma}{Lemma}
\newtheorem{definition}{Definition}

\newtheorem{assumption}{Assumption}

\newcommand*{\QED}{\hfill\ensuremath{\square}}

 \def\old#1{}    

\def\RED{{\mbox{\rm\tiny R-ED}}}

\def\beq{\begin{equation}}
\def\eeq{\end{equation}}
\def\bea{\begin{eqnarray}}
\def\eea{\end{eqnarray}}
\def\ba{\begin{array}}
\def\ea{\end{array}}

\def\bitem{\begin{itemize}}
\def\eitem{\end{itemize}}
\def\ben{\begin{enumerate}}
\def\een{\end{enumerate}}

\def\ie{{\it i.e.,\ \/}}

\definecolor{bgrd}{rgb}{1,1,1}
\definecolor{gray}{rgb}{0.5,0.5,0.5}
\definecolor{dkr}{rgb}{0.7,0.1,0.2}
\definecolor{dkb}{rgb}{0.1,0.1,0.8}

\makeatletter
\newdimen{\captionwidth}
\long\def\@makecaption#1#2{%
\captionwidth .9\hsize
\vskip 10pt%
\setbox\@tempboxa\hbox{#1: #2}%
  \ifdim \wd\@tempboxa >\captionwidth%
    \setbox\@tempboxa\hbox{#1:\hspace*{.5em}}%
    \hfil\parbox{\captionwidth}{\raggedright\hangindent \wd\@tempboxa%
    \hangafter=1\unhbox\@tempboxa#2}\hfill%
  \else\centerline{\box\@tempboxa}%
  \fi
}
\makeatother
\def\scalefig#1{\epsfxsize #1\textwidth}

\def\edoc{\end{document}}

\def\T{\mbox{\tiny T}}

\def\zetabf{\hbox{\boldmath$\zeta$\unboldmath}}

\def\lambdabf{\hbox{\boldmath$\lambda$\unboldmath}}
\def\mubf{\hbox{\boldmath$\mu$\unboldmath}}

\def\xibf{\hbox{\boldmath$\xi$\unboldmath}}
\def\pibf{\hbox{\boldmath$\pi$\unboldmath}}
\def\rhobf{\hbox{\boldmath$\rho$\unboldmath}}

\def\phibf{\hbox{\boldmath$\phi$\unboldmath}}
\def\chibf{\hbox{\boldmath$\chi$\unboldmath}}
\def\psibf{\hbox{\boldmath$\psi$\unboldmath}}

\def\Deltabf{\hbox{$\bf \Delta$}}

\def\thetabf{{\mbox{\boldmath$\theta$\unboldmath}}}

\def\cbf{{\bm c}}
\def\dbf{{\bm d}}
\def\ebf{{\bm e}}

\def\gbf{{\bm g}}

\def\pbf{{\bm p}}
\def\qbf{{\bm q}}
\def\rbf{{\bm r}}
\def\sbf{{\bm s}}

\def\rbf{{\bm r}}

\def\Abf{{\bm A}}
\def\Bbf{{\bm B}}

\def\Ebf{{\bm E}}

\def\Gbf{{\bm G}}

\def\Ec{{\cal E}}

\def\Gc{{\cal G}}
\def\Hc{{\cal H}}

\begin{document}

\title{Wholesale Market Participation of Storage  with State-of-Charge Dependent Bids}
\author{Cong Chen,~\IEEEmembership{Student Member,~IEEE}
and~Lang~Tong,~\IEEEmembership{Fellow,~IEEE}
\thanks{\scriptsize
Cong Chen (\url{cc2662@cornell.edu}) and Lang Tong (\url{lt35@cornell.edu}) are with the Cornell University, Ithaca, NY 14853, USA.  }
\thanks{\scriptsize The work of L. Tong and C. Chen is supported in part by the National Science Foundation under Award 2218110 and 1932501.}}
\maketitle

\begin{abstract}
Wholesale market participation of storage with state-of-charge (SoC) dependent bids results in a non-convex cost in a multi-interval economic dispatch, which requires a mixed-integer linear program in the market clearing.  We show that the economic dispatch can be convexified to the standard linear program when the SoC-dependent bid satisfies the equal decremental-cost ratio (EDCR) condition.  Such EDCR bids are shown to support individual rationalities of all market participants in both the day-ahead multi-interval economic dispatch under locational marginal pricing and the rolling-window look-ahead dispatch under temporal-locational marginal pricing in the real-time market.  A numerical example is presented to demonstrate a higher profit margin with an SoC-dependent bid over that from an SoC-independent bid.
\end{abstract}


\section{Introduction} \label{sec:intro}
There have been recent proposals that allow storage participants in the wholesale electricity market to submit state-of-charge (SoC) dependent offers and bids  \cite{CAISO_SOCdependent:22}. Such bids incorporate SoC-dependent operation \cite{ecker14SoCDegradation} and opportunity costs  \cite{ZhengXu22socAribitrage} of merchant storage participants into bidding parameters submitted to a bid-based market clearing process. In that way, the central market clearing can produce an economic dispatch program that schedules the battery SoC within a range favorable to the battery's health and the storage's ability to capture future profit opportunities under uncertainty. 

However, SoC-dependent bids and offers result in a non-convex optimization for the multi-interval dispatch in the electricity market, causing computationally expensive market clearing processes, especially when we have large amounts of storage participating in the market. Such nonconvexity also distorts the current locational marginal pricing (LMP)  signals, requiring out-of-the-market uplifts in the day-ahead and real-time markets to support the dispatch-following incentive. 

The LMP signal is also distorted by the inconsistency of rolling-window look-ahead dispatch in the real-time market \cite{hogan2020Multi-intervalPricing}. Recently, the real-time electricity market in California ISO (CAISO) has seen frequent out-of-merit storage dispatches in the rolling-window dispatch \cite{CAISO21MIO} that requires the discriminative out-of-the-market uplift payments to ensure storage participants follow the operator's dispatch signals. 

This paper aims to uncover causes of nonconvexity from SoC-dependent bids and offer simple remedies that convexify the multi-interval economic dispatch to conform with standard market clearing and pricing processes. 
\subsection{Related work}
The cost structure of storage is explored by many literature and experimental results, which can be categorized into (i) depth-of-discharge (DoD) dependent cost \cite{ecker14SoCDegradation, Shi18DoDstorageRegulation, Xu22MSRstorageCost, HeKar21TPSdegradation} , (ii) SoC-dependent degradation cost \cite{ecker14SoCDegradation, Xu22MSRstorageCost, Amine01JPSbattergAging, Vetter05JPSAging, Safari10JElectrochemSocBatteryAging}  and (iii) SoC-dependent opportunity cost\cite{ZhengXu22socAribitrage, Xu20CDCstorageValue}. Incorporating these costs in dispatch decisions is directly or indirectly related to the SoC operating range of the storage, and has the potential to reduce storage operating costs and bring economic benefits to merchant storage participants. 
 
To include SoC-dependent cost into the market clearing problem, some of the literature adopted SoC-dependent weights with a mixed integer program when computing the throughput of storage \cite{LiuWangXuGuo17IETdegradation, ZhaoQiu18robustESloss, Chen19PriciseES}, making the storage operation cost SoC-dependent. But these authors are not intended for the deregulated electricity market with a bid-based market clearing process. Most recently, CAISO has initiated discussions about allowing SoC-dependent bids for storage, which is a piecewise linear function between the storage bids/offers and the SoC \cite{CAISO_SOCdependent:22}. Furthermore,  a market clearing mechanism considering such SoC-dependent bids is formulated into a mixed integer program in \cite{ZhengQinWuMurtagughXu22workingSOC}. Note that, all existing market clearing solutions to SoC dependent bids rely on mixed integer programs, making the market clearing process computationally expensive for large-scale practical implementations. The nonconvexity of the market clearing problem brought by the SoC-dependent bid also presents pricing challenges, resulting in out-of-merit dispatch and the need for out-of-the-market settlements in both one-shot and rolling-window dispatch.

To the best of our knowledge, our earlier work  \cite{ChenTong2022convexifying} is the first to convexify the market clearing problem with SoC-dependent bids, thus removing the necessity of integer variables, although it doesn't explore the influence of SoC-dependent bids on the pricing schemes in the one-shot and rolling-window dispatch. 

This paper is also related to literature about pricing storage operation in the multi-interval dispatch under uncertainty. Many researchers have analyzed the dispatch-following incentive issue of LMP caused by the inconsistency of rolling-window dispatch, and propose new pricing schemes for generators \cite{hogan2020Multi-intervalPricing, Hua&etal:19TPS,Zhao&Zheng&Litvinov:19TPS, Guo&Chen&Tong:21TPS, Chen&Guo&Tong:20TPS} and storage \cite{Zhao&Zheng&Litvinov:19TPS, ChenTong22EPSRstorage}. Our earlier result shows that, in the rolling-window dispatch, there's no uniform price that can fully support dispatch-following incentives without out-of-the-market uplifts\cite{Guo&Chen&Tong:21TPS, ChenTong22EPSRstorage}. And our idea of generalizing LMP to the nonuniform temporal locational marginal pricing (TLMP) for storage  \cite{ChenTong22EPSRstorage}, can support the dispatch-following incentive without uplifts for merchant storage participants. Here, we further extend our earlier results about TLMP with SoC-independent bids \cite{ChenTong22EPSRstorage} to the rolling-window dispatch with SoC-dependent bids.

\subsection{Summary of results}

This paper focuses on the convexification and pricing of the SoC-dependent multi-interval market clearing problem. The main contribution of this work is threefold.  

First, we provide insights into non-convexity introduced by SoC-dependent cost and the rationale for imposing the so-called equal decremental-cost ratio (EDCR) constraint proposed in \cite{ChenTong2022convexifying}. We also provide a new cost expression (c.f. Theorem~\ref{thm:Eq}) under EDCR that is critical in pricing storage with SoC-dependent costs in rolling-window look-ahead scheduling.

Second, we establish that the standard LMP guarantees dispatch-following incentives in the one-shot multi-interval economic dispatch typically used in the day-ahead market with bids satisfying the EDCR condition. We also provide a numerical example demonstrating the impact of EDCR SoC-dependent bids\footnote{EDCR SoC-dependent bids refers to the SoC-dependent bid satisfying the EDCR condition.} on the profit, the millage, and the profit margin of storage participants. 

Finally, we show that  EDCR bids priced under TLMP guarantee dispatch-following incentives for rolling-window dispatch, independent of forecasting errors. In particular, the rolling-window dispatch of storage achieves the maximum individual profit for the storage given the realized (ex-post) price over the entire scheduling horizon. 

This paper is organized as follows. We first introduce the SoC-dependent cost structure in Sec.~\ref{sec:model}.  Sec.~\ref{sec:DAM} provides insights into the nonconvexity induced by SoC-dependent costs and the rationale that the EDCR condition convexifies the dispatch optimization. In Sec.~\ref{sec:RTM}, we introduce the rolling-window dispatch with the SoC-dependent bid and the effects of TLMP.  A numerical example illustrating the benefits of SoC-dependent bids is given in Sec.~\ref{sec:NE},

\section{SoC-dependent bid and stage cost models} \label{sec:model}
In this section, we will introduce the SoC dependency of storage operation cost, and then explain the SoC-dependent bid together with the single-stage cost models.

\subsection{SoC-dependent battery degradation cost}
\begin{figure}[h]
\center
\begin{psfrags}
\scalefig{0.3}\epsfbox{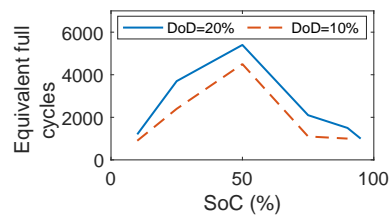}
\end{psfrags}
\vspace{-1.5em}\caption{\scriptsize SoC-dependent degradation cost of storage. }
\label{fig:SOCCyc}
\end{figure}

The battery degradation cost is shown to be SoC-dependent by many lab degradation test results like \cite{ecker14SoCDegradation, Amine01JPSbattergAging, Vetter05JPSAging, Safari10JElectrochemSocBatteryAging}.  A general relation between SoC and battery lifetime degradation is shown in Fig.~\ref{fig:SOCCyc}, plotted based on data from \cite{ecker14SoCDegradation}. It can be observed that storage has more equivalent full cycles when operating in the middle SoC range. This means storage degradation cost is lower in the middle SoC range than that approaching SoC operating boundaries. In the current electricity market, merchant storage participants can only participate with a SoC-independent bid \cite{CAISO20ESRbid}. This means storage needs to do approximation over the true SoC-dependent cost, or keep updating the bid-in parameters for costs and physical limits based on its SoC. 

\subsection{SoC-dependent bid and cost models}
\begin{figure}[h]
\center
\begin{psfrags}
\scalefig{0.35}\epsfbox{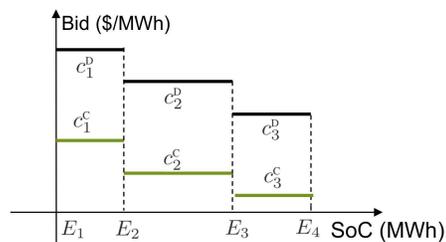}
\end{psfrags}
\vspace{-1.5em}\caption{\scriptsize  SoC-dependent bid. }
\label{fig:SOC_D_Cost}
\end{figure}

To reduce the barrier to the economic storage operation, recent proposals  \cite{CAISO_SOCdependent:22} have allowed merchant storage participants in wholesale electricity to submit SoC-dependent offers and bids. A standard piecewise-linear SoC-dependent bid model is shown in Fig.~\ref{fig:SOC_D_Cost}. We partition the SoC axis into $K$ consecutive segments, within each segment $\Ec_k=[E_k,E_{k+1}]$, a pair of bid-in cost/benefit parameters  $(c_k^{\mbox{\tiny C}},c_k^{\mbox{\tiny D}})$ is defined.  This SoC-dependent bid is composed of bid-in parameters $\mathbf{c}^{\mbox{\tiny C}}:=(c_k^{\mbox{\tiny C}})$, $\mathbf{c}^{\mbox{\tiny D}}:=(c_k^{\mbox{\tiny D}})$ and $\mathbf{E}:= (E_k)$. For simplicity, storage index and storage ramping costs are ignored in this section.

Additionally, we introduce Assumption~\ref{assume:single} for the SoC-dependent bids based on two considerations. Firstly, for the longevity of the battery and the ability to capture profit opportunities, it is more costly to discharge when the SoC is low, and the benefit of charging is less when the SoC is high. Therefore, typical  bid-in discharge costs $(c^{\mbox{\tiny D}}_{k})$ and charging benefits $(c^{\mbox{\tiny C}}_{k})$ are monotonically decreasing. Secondly, the storage participant is willing to discharge only if the selling price is higher than the buying price. Hence, the storage participant's willingness to sell by discharge (adjusted to discharging efficiency) must be higher than its willingness to purchase (adjusted to charging efficiency),  \ie $c^{\mbox{\tiny D}}_{k}\eta^{\mbox{\tiny D}} > c^{\mbox{\tiny C}}_{k}/\eta^{\mbox{\tiny C}}, \forall k$, with charging/discharging efficiencies  $\eta^{\mbox{\tiny C}},\eta^{\mbox{\tiny D}} \in (0,1]$.  

\begin{assumption}[] \label{assume:single} The SoC-dependent cost/benefit parameters $\{(c_k^{\mbox{\rm \tiny C}},c_k^{\mbox{\rm \tiny D}}), \eta^{\mbox{\rm \tiny C}}, \eta^{\mbox{\rm \tiny D}}\}$ satisfy the following monotonicity conditions $\forall k=1,\cdots, K-1$:
\[
\left\{\begin{array}{l}
c_k^{\mbox{\rm \tiny C}} \ge c_{k+1}^{\mbox{\rm \tiny C}}\\
c_k^{\mbox{\rm \tiny D}} \ge c^{\mbox{\rm \tiny D}}_{k+1}\\
\end{array}\right.~~{\rm and}~~ c^{\mbox{\rm \tiny C}}_{1}/\eta^{\mbox{\rm \tiny C}} < c^{\mbox{\rm \tiny D}}_{\mbox {\rm \tiny K}}\eta^{\mbox{\rm \tiny D}}.\]
\end{assumption}
 
As for the physical operation parameters of storage, we adopt the standard imperfect storage model.  In the scheduling interval $t$, let  $e_t$ be the storage SoC, $g_t^{\mbox{\tiny C}}$ the charging power, and $g_t^{\mbox{\tiny D}}$ the discharging power, respectively. The storage SoC evolves according to 
\beq
\begin{array}{lcl}
e_{t+1}=e_{t} +g^{\mbox{\tiny C}}_{t}\eta^{\mbox{\tiny C}}-g^{\mbox{\tiny D}}_{t}/\eta^{\mbox{\tiny D}},~~ g^{\mbox{\tiny C}}_{t}g^{\mbox{\tiny D}}_{t}=0,
\end{array}
\eeq

SoC-dependent bids and offers induce SoC-dependent scheduling costs involving the ({\it ex-ante}) SoC $e_t$ in scheduling stage $t$ before dispatch and the ({\it ex-post}) SoC $e_{t+1}$ that may be in a different SoC partitioned segment.  Specifically, the stage cost in interval $t$ is given by 
\beq\label{eq:StageCost}
f(g^{\mbox{\tiny C}}_t, g^{\mbox{\tiny D}}_t;e_t)= f^{\mbox{\tiny D}}(e_t,g_t^{\mbox{\tiny D}})-f^{\mbox{\tiny C}}(e_t,g_t^{\mbox{\tiny C}}),
\eeq
where $f^{\mbox{\tiny C}}(\cdot)$ and $f^{\mbox{\tiny D}}(\cdot)$ are charging and discharging costs. For every  $e_t \in \Ec_m, e_{t+1}=e_{t} +g^{\mbox{\tiny C}}_{t}\eta^{\mbox{\tiny C}}-g^{\mbox{\tiny D}}_{t}/\eta^{\mbox{\tiny D}} \in \Ec_{n}$,  $f^{\mbox{\tiny C}}(\cdot)$ and $f^{\mbox{\tiny D}}(\cdot)$ are respectively defined as follows:  
\[f^{\mbox{\tiny C}}(e_t, g_t^{\mbox{\tiny C}}) := 
\left\{\begin{array}{ll}
 g^{\mbox{\tiny C}}_tc^{\mbox{\tiny C}}_{n}   & m=n\\
  g^{\mbox{\tiny C}}_tc^{\mbox{\tiny C}}_{n} +\sum_{k=m}^{n-1}\frac{\Delta c^{\mbox{\tiny C}}_{k}}{\eta^{\mbox{\tiny C}}}(E_{k+1}-e_t) & n>m\\
0 & \mbox{otherwise}
\end{array}\right.\]
\[f^{\mbox{\tiny D}}(e_t, g_t^{\mbox{\tiny D}}) := 
\left\{\begin{array}{ll}
 g^{\mbox{\tiny D}}_tc^{\mbox{\tiny D}}_{n}   &  m=n\\
 g^{\mbox{\tiny D}}_tc^{\mbox{\tiny D}}_{n} +\sum_{k=n+1}^{m}\eta^{\mbox{\tiny D}}\Delta c^{\mbox{\tiny D}}_{k}(E_{k}-e_t) & n<m\\
0 & \mbox{otherwise}
\end{array}\right.
\]
with $\Delta c^{\mbox{\tiny C}}_{k}:=c^{\mbox{\tiny C}}_{k}-c^{\mbox{\tiny C}}_{k+1}$ and $\Delta c^{\mbox{\tiny D}}_{k}:=c^{\mbox{\tiny D}}_{k}-c^{\mbox{\tiny D}}_{k+1}$. The stage cost above is first explained in \cite{ChenTong2022convexifying} together with an example illustrating the computation of $f^{\mbox{\tiny C}}(e_t, g_t^{\mbox{\tiny C}})$ from the perspective of lebesgue integral.  Note that the stage cost  $f(g_t^{\mbox{\tiny C}},g_t^{\mbox{\tiny D}}; e_t)$ is non-convex, although it is convex if given $e_t$.

\section{One-shot dispatch and pricing} \label{sec:DAM}
In this section, we first explain the one-shot multi-interval economic dispatch model adopted in the day-ahead electricity market and then propose a sufficient condition to convexify the market clearing problem for the one-shot dispatch. A toy example is illustrated here to give intuitive virtualization of the convexification procedure. Furthermore, under one-shot LMP, the individual rationality and truthful-bidding incentive for storage with the SoC-dependent bid is analyzed.


 \subsection{Multi-interval economic dispatch and one-shot LMP}
We consider a {\em bid-based electricity market}  involving one inelastic demand,  $N$ storage units, and a system (market) operator\footnote{Network constraints are ignored here for simplicity, and our analysis can be easily extended to the case with network constraints.}.  And we model all market participants with the generalized storage model \cite{ChenTong22EPSRstorage}. For example, the generalized storage model for elastic demand has the charge power representing the demand, and the SoC representing the accumulated consumption. The physical limit is zero for discharge power and infinite for SoC. 

The scheduling period of storage involves $T$ unit-length  intervals $\Hc=\{1,\cdots, T\}$, where interval $t$ covers the time interval $[t,t+1)$.  Typically,   $T$  is the number of intervals in a day. Denote the charge power vector of storage indexed by $i$ by $\gbf^{\mbox{\tiny C}}_i:=(g^{\mbox{\tiny C}}_{i1},\cdots, g^{\mbox{\tiny C}}_{iT})$ (and,  similarly $\gbf^{\mbox{\tiny D}}_i$). Given the initial SoC $e_{i1}=s_i$ and $\Gbf:=\{\gbf^{\mbox{\tiny C}}_i, \gbf^{\mbox{\tiny D}}_i\}$ , the total cost of $N$ storage over multi-interval $\Hc$ is  \beq\begin{array}{lrl}\label{eq:Totalcost}
  \bar{F}(\Gbf, \sbf) &= & \sum_{i=1}^N F_i(\gbf^{\mbox{\tiny C}}_i, \gbf^{\mbox{\tiny D}}_i;s_i)\\
 & =&\sum_{i=1}^N \sum_{t=1}^{T}f_i(g^{\mbox{\tiny C}}_{it}, g^{\mbox{\tiny D}}_{it};e_{it}).
 \end{array}\eeq
 
Given the load forecast $(\hat{d}_t)$, the  $T$-interval economic dispatch minimizes the total system operation costs is 
\beq \label{eq:NONCVX}
\begin{array}{lrl}
\Gc: &\underset{\{(g_{it}^{\mbox{\tiny C}}, g_{it}^{\mbox{\tiny D}}, e_{it})\}}{\rm minimize} &  \bar{F}(\Gbf, \sbf)=\sum_{i=1}^{N} F_i(\gbf^{\mbox{\tiny C}}_i, \gbf^{\mbox{\tiny D}}_i;s_i)\\& s.t.& \forall t\in \Hc, \forall i\in \{1,...,N\}\\
&\lambda_{t}:& \sum_{i=1}^{N}(g^{\mbox{\tiny D}}_{it}-g^{\mbox{\tiny C}}_{it})=\hat{d}_{t},\\
&(\underline{\mu}^{\mbox{\tiny C}}_{it},\bar{\mu}^{\mbox{\tiny C}}_{it}):& -\underline{r}^{\mbox{\tiny C}}_i \le g^{\mbox{\tiny C}}_{it}-g^{\mbox{\tiny C}}_{i(t-1)} \le \bar{r}^{\mbox{\tiny C}}_i,\\
&(\underline{\mu}^{\mbox{\tiny D}}_{it},\bar{\mu}^{\mbox{\tiny D}}_{it}):& -\underline{r}^{\mbox{\tiny D}}_i  \le g^{\mbox{\tiny D}}_{it}-g^{\mbox{\tiny D}}_{i(t-1)}\le  \bar{r}^{\mbox{\tiny D}}_i,\\
&\phi_{it}:& e_{it}+g^{\mbox{\tiny C}}_{it}\eta^{\mbox{\tiny C}}-g^{\mbox{\tiny D}}_{it}/\eta^{\mbox{\tiny D}}=e_{i(t+1)},\\
&&\underline{e}_i \le e_{i(t+1)}\le \bar{e}_i,e_{i1}=s_i,\\
&& 0 \le g^{\mbox{\tiny C}}_{it}\le \bar{g}^{\mbox{\tiny C}}_i,\\
&& 0 \le g^{\mbox{\tiny D}}_{it}\le \bar{g}^{\mbox{\tiny D}}_i,
\end{array}
\eeq
where operation constraints sequentially included are power balance constraints, charging \& discharging ramp limits, SoC state transition constraints, SoC limits, and charging \& discharging capacity limits. Dual variables most relevant in defining prices are SoC shadow prices $\phi_{it}$, ramping shadow prices $(\underline{\mu}^{\mbox{\tiny C}}_i,\bar{\mu}^{\mbox{\tiny C}}_i, \underline{\mu}^{\mbox{\tiny D}}_i,\bar{\mu}^{\mbox{\tiny D}}_i)$, and power balance shadow prices $\lambda_{it}$. 

Note that we ignore the non-convex constraint $g^{\mbox{\tiny C}}_{t}g^{\mbox{\tiny D}}_{t}=0, \forall t$  that prevents simultaneous charging/discharging decisions  in (\ref{eq:NONCVX}). Following references \cite{Li18CSEEstorage, ChenBaldick21TPSstorageSCUCBinary},  the following lemma first proposed in   \cite{ChenTong2022convexifying} shows that such a relaxation can be justified for cases with non-negative LMPs. 

\begin{lemma}[] \label{lemma:bidSpread}
Under Assumption~\ref{assume:single}  and non-negative LMPs, the optimal solution of (\ref{eq:NONCVX}) satisfies $g^{\mbox{\tiny C}*}_{it}g^{\mbox{\tiny D}*}_{it}=0, \forall i, t$. 
\end{lemma}

Although the constraints of (\ref{eq:NONCVX}) are convex, the objective function is not because the stage-cost function $F_i(\cdot)$ is non-convex in a multi-interval dispatch. Therefore, the SoC-dependent bid causes nonconvexity in the multi-interval economic dispatch. And pricing non-convex multi-interval dispatch becomes nontrivial. However, for some choices of bidding parameters, the objective is convex, for which an example is shown in Fig.~\ref{fig:SOCdependent} (top right). In this case, the  {\em one-shot locational  marginal price\footnote{We retain the LMP terminology even though the model considered here does not involve a network.}}  (LMP for short) is a uniform price $(\pi^{\mbox{\tiny LMP}}_t)$ with $\pi^{\mbox{\tiny LMP}}_t$ defined by  the marginal energy providing cost with respect to the demand in interval $t$. In particular, we have, by the envelope theorem,
\beq \label{eq:RWLMP}
\pi^{\mbox{\rm\tiny LMP}}_{t} := \frac{\partial}{\partial d_t} \bar{F}(\Gbf^*, \sbf) = \lambda_t^*,
\eeq
where $\bar{F}(\Gbf^*, \sbf) $ is the objective (\ref{eq:Totalcost}) at the optimal solution. 

\subsection{A sufficient condition convexifying multi-interval dispatch}
Theorem~\ref{thm:Eq} below gives a condition on bid-in cost parameters that convexify the market clearing optimization (\ref{eq:NONCVX}). For simplicity, storage index $i$ is omitted here.
\begin{theorem}[] \label{thm:Eq}
If a storage participant's bid-in parameters satisfy the equal decremental-cost ratio (EDCR) condition,
\beq\label{eq:EDCR}
\frac{c^{\mbox{\rm \tiny C}}_k-c^{\mbox{\rm \tiny C}}_{k-1}}{c^{\mbox{\rm \tiny D}}_k-c^{\mbox{\rm \tiny D}}_{k-1}}=\eta^{\mbox{\rm \tiny C}}\eta^{\mbox{\rm \tiny D}}, \forall k, 
\eeq
under Assumption~\ref{assume:single}  and non-negative LMPs, the total cost of a storage dispatch over multi-interval for the given initial SoC $e_1=s$ is a piecewise linear convex function of $\gbf^{\mbox{\tiny C}}:=(g_t^{\mbox{\rm \tiny C}})$ and $\gbf^{\mbox{\rm \tiny D}} = (g_t^{\mbox{\rm \tiny D}})$ given by 
\beq
\begin{array}{lrl}\label{eq:ES_cost}
F(\gbf^{\mbox{\rm \tiny C}}, \gbf^{\mbox{\rm \tiny D}};s)=\underset{j\in\{1,...,K\}}{\rm max}\{\alpha_j(s)-c^{\mbox{\rm \tiny C}}_{j}\mathbf{1}^\intercal \gbf^{\mbox{\rm \tiny C}}+c^{\mbox{\rm \tiny D}}_{j}\mathbf{1}^\intercal \gbf^{\mbox{\rm \tiny D}}\},\\
~~~~~~~~~~~~~~=-c^{\mbox{\tiny C}}_{n}\mathbf{1}^\intercal \mathbf{g}^{\mbox{\tiny C}}+c^{\mbox{\tiny D}}_{n}\mathbf{1}^\intercal \mathbf{g}^{\mbox{\tiny D}}\\
~~~~~~~~~~~~~~~+\begin{cases}
\sum_{k=m}^{n-1}\frac{-\Delta c^{\mbox{\tiny C}}_{k}}{\eta^{\mbox{\tiny C}}}(E_{k+1}-s), & n>m \\
0, & m=n  \\
\sum_{k=n+1}^{m}\eta^{\mbox{\tiny D}}\Delta c^{\mbox{\tiny D}}_{k-1}(E_{k}-s), & n<m
\end{cases},
\end{array}
\eeq
where $\alpha_j(s):=-\sum_{k=1}^{j-1}\frac{\Delta c^{\mbox{\rm \tiny C}}_k(E_{k+1}-E_1)}{\eta^{\mbox{\rm \tiny C}}}-\frac{c^{\mbox{\rm \tiny C}}_{j}(s-E_1)}{\eta^{\mbox{\rm \tiny C}}}+h^{\mbox{\rm \tiny C}}(s)$, $h^{\mbox{\rm \tiny C}}(s):=\sum_{i=1}^K\mathbb{I}\{s\in \Ec_i\}(\frac{c^{\mbox{\rm \tiny C}}_i(s-E_1)}{\eta^{\mbox{\rm \tiny C}}}+\sum_{k=1}^{i-1}\frac{\Delta c^{\mbox{\rm \tiny C}}_k (E_{k+1}-E_1)}{\eta^{\mbox{\tiny C}}})$\footnote{$\mathbb{I}\{s \in \Ec_i\}$ is indicator function, which equals to 1 when $s \in \Ec_i$.}. $m$ and $n$ are respectively indexes for SoC partitioned segments that the initial and the end-state SoC respectively fall into, \ie $e_{\mbox{\rm \tiny 1}}=s\in \Ec_{m}$ and $e_{\mbox{\rm \tiny T+1}}\in \Ec_{n}$.
\end{theorem}
The EDCR condition is first proposed in \cite{ChenTong2022convexifying} with detailed proof, and here we provide an alternative new equivalent formulation of  the multi-interval storage cost, which indicates the multi-interval cost is linear given the initial and the end-state SoC. This is critical in pricing the rolling-window dispatch in Sec.~\ref{sec:RTM} with the SoC-dependent bid considered.
\subsection{Convexification Insights of the EDCR condition}\label{sec:ConvexAffine}
Knowing that a function is non-convex if the composition of the function with an affine function is non-convex \cite{Boyd:04CVX}, we use such a composition to demonstrate the insight of the nonconvexity and convexification in the following example.
\begin{figure}[h]
\center
\begin{psfrags}
\scalefig{0.5}\epsfbox{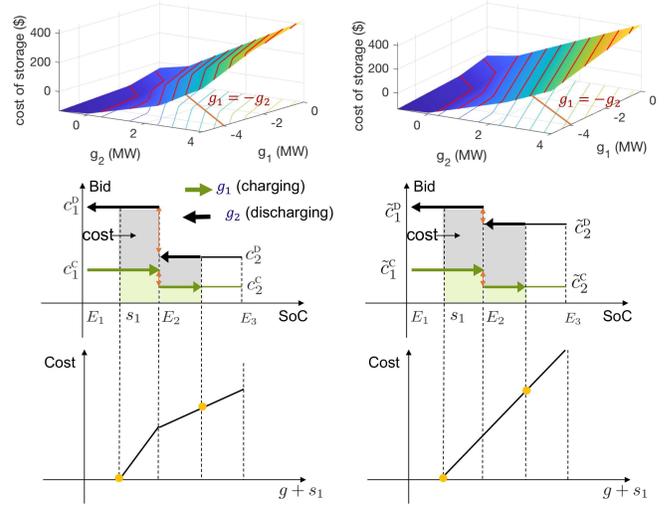}
\end{psfrags}
\vspace{-0.5em}\caption{\scriptsize Top: SoC-dependent cost of storage in the 2-interval dispatch (left:  non-convex true cost; right: convex EDCR cost. Middle: SoC-dependent bid (left: true SoC-dependent marginal cost; right: EDSR  bid. Bottom: the composition of storage cost function with the affine function $g_1=-g_2$ (left: non-convex true cost; right: convex EDCR cost). }
\label{fig:SOCdependent}
\end{figure}

Consider ideal storage in a 2-interval dispatch with true non-convex multi-interval storage cost shown in the top left of Fig.~\ref{fig:SOCdependent}, and convexified storage cost function based on the EDCR bid shown in Fig~\ref{fig:SOCdependent} (top right). The x-axis notation represents the net-producing power of storage, defined by $g_t:=g^{\mbox{\rm \tiny D}}_{t}-g^{\mbox{\rm \tiny C}}_{t}, \forall t=1,2$.  Detailed parameters of this example are explained in Sec.~\ref{sec:NE}.  Figures related to the true SoC-dependent marginal cost are shown in the left column of Fig.~\ref{fig:SOCdependent}, and figures related to the EDCR SoC-dependent bid are shown in the right column. 

The composition of storage cost function with the affine function  $g_1=-g_2$ is shown in the bottom of Fig.~\ref{fig:SOCdependent} where $g_1=g$. Such affine function restricts that the storage will always loop back to the initial SoC in this 2-interval dispatch. The slope of Fig.~\ref{fig:SOCdependent} (bottom) is $c^{\mbox{\rm \tiny D}}_1-c^{\mbox{\rm \tiny C}}_{1}$ when $g+s_1\in \Ec_1$, and  $c^{\mbox{\rm \tiny D}}_2-c^{\mbox{\rm \tiny C}}_{2}$ when $g+s_1\in \Ec_2$. With true SoC-dependent bidding parameters $(\cbf^{\mbox{\rm \tiny C}}, \cbf^{\mbox{\rm \tiny D}}, \Ebf)$ shown in Fig.~\ref{fig:SOCdependent} (middle left), the EDCR condition  (\ref{eq:EDCR}) is not satisfied, and the nonconvexity is observed at the breakpoint  $E_2$ in Fig.~\ref{fig:SOCdependent} (bottom left). But with the EDCR bids $(\tilde{\cbf}^{\mbox{\rm \tiny C}}, \tilde{\cbf}^{\mbox{\rm \tiny D}})$\footnote{The EDCR conditions in (\ref{eq:EDCR}) decreased to $\tilde{c}^{\mbox{\rm \tiny C}}_k-\tilde{c}^{\mbox{\rm \tiny C}}_{k-1}=\tilde{c}^{\mbox{\rm \tiny D}}_k-\tilde{c}^{\mbox{\rm \tiny D}}_{k-1}, \forall k$, for this ideal storage, \ie $\eta^{\mbox{\rm \tiny D}}=\eta^{\mbox{\rm \tiny C}}=1$. } shown in Fig.~\ref{fig:SOCdependent} (middle right), we have  $\tilde{c}^{\mbox{\rm \tiny D}}_{1}-\tilde{c}^{\mbox{\rm \tiny C}}_{1}=\tilde{c}^{\mbox{\rm \tiny D}}_2-\tilde{c}^{\mbox{\rm \tiny C}}_2$. So, the composition of the storage cost function with the affine function $g_1=-g_2$ is linear, as is illustrated in Fig.~\ref{fig:SOCdependent} (bottom right).


\subsection{Individual rationality and truthful-bidding under LMP,}
Apart from the market clearing conditions restricted by the power balance constraint, the individual rationality defined below is the key for a general equilibrium price \cite[p. 547]{Mas-Colell&Winston&Green:95book}. 
\begin{definition}[Individual rationality for storage\footnote{Storage index $i$ is omitted here for simplicity.}] \label{def:IR}
We say vectors of multi-interval price and dispatch $(\pibf, \gbf^{\mbox{\rm \tiny C}}, \gbf^{\mbox{\rm \tiny D}})$ support individual rationality for storage, if the dispatch is the solution to the individual profit maximization given by:
\beq \label{eq:Q}
\begin{array}{lcl}
Q(\pibf)=& \underset{\{\pbf^{\mbox{\rm \tiny D}},\pbf^{\mbox{\rm \tiny C}},\qbf\}}{\rm maximize} & \big(
 \pibf^{\intercal}\pbf^{\mbox{\rm \tiny D}}-\pibf^{\intercal}\pbf^{\mbox{\rm \tiny C}}-F(\pbf^{\mbox{\rm \tiny C}}, \pbf^{\mbox{\rm \tiny D}};s)\big)\\
& {\rm s.t.}& \eta^{\mbox{\rm \tiny C}}\pbf^{\mbox{\rm \tiny C}}-\pbf^{\mbox{\rm \tiny D}}/\eta^{\mbox{\rm \tiny D}}=\Abf\qbf, \\
&& -\underline{\rbf}^{\mbox{\rm \tiny D}}\le \Bbf \pbf^{\mbox{\rm \tiny D}} \le \bar{\rbf}^{\mbox{\rm \tiny D}}, \\
&& -\underline{\rbf}^{\mbox{\rm \tiny C}}\le \Bbf \pbf^{\mbox{\rm \tiny C}} \le \bar{\rbf}^{\mbox{\rm \tiny C}}, \\
&& \underline{\ebf}\le \qbf \le \bar{\ebf},  q_1=s,\\
&&{\bf 0}\leq \pbf^{\mbox{\rm \tiny D}} \leq\bar{\gbf}^{\mbox{\rm \tiny D}},\\
&&{\bf 0}\leq \pbf^{\mbox{\rm \tiny C}} \leq \bar{\gbf}^{\mbox{\rm \tiny C}},
\end{array} \hfill
\eeq
where  $\pbf^{\mbox{\rm \tiny D}},\pbf^{\mbox{\rm \tiny C}},\qbf$ are, respectively, decision variables for  discharging power, charging power and SoC. The SoC evolution is defined by matrix $\Abf \in \mathbb{R}^{\rm T \times (T+1)}$ with most zero elements except for $A_{ii}=-1$ and $A_{i(i+1)}=1, \forall i \in \{1,..,T\}$. The ramping constraints are defined by lower bidiagonal matrix $ \Bbf \in \mathbb{R}^{\rm T \times T}$ with $1$ as diagonals and $-1$ left next to diagonals.
\end{definition}

In the context of analyzing dispatch-following incentives, we are interested in whether price signal $\pibf$ and dispatch  $\Gbf$ satisfy the Individual rationality  condition for all market participants. It turns out that, in the absence of forecasting error, the one-shot LMP supports the one-shot economic dispatch as stated  in Theorem~\ref{thm:LOC_MIO}. 

\begin{theorem}[Individual rationality and truthful-bidding incentive under one-shot LMP] \label{thm:LOC_MIO} When the EDCR condition and Assumption~\ref{assume:single}  are satisfied, the one-shot LMP and dispatch  $(\pibf^{\mbox{\rm\tiny LMP}}, \gbf_i^{\mbox{\rm \tiny D}*},\gbf_i^{ \mbox{\rm \tiny C}*})$ over $\Hc$  computed from (\ref{eq:NONCVX}) can support individual rationality for storage $i$. And it is optimal for a price-taking storage to bid truthfully with its marginal costs of charging and discharging. 
\end{theorem}
{\em Proof:}   See Appendix for the proof.

This result is analogous to the well-known property of LMP \cite{Wu&Variaya&Spiller&Oren:96JRE}. The proof of Theorem~\ref{thm:LOC_MIO} replies on the convexity of (\ref{eq:NONCVX}) when the EDCR condition is satisfied. 


\section{Rolling-window dispatch and pricing} \label{sec:RTM}
In this section, we first introduce the rolling-window dispatch, and the definition of lost opportunity cost. Then, we extend our previous results of rolling-window temporal location marginal pricing (R-TLMP) with SoC-independent bids \cite{ChenTong22EPSRstorage} to the case with SoC-dependent bids and end-state SoC control.

\subsection{Rolling-window settlement and pricing rules}
The rolling-window economic dispatch policy relies on solving a series of one-shot economic dispatch $\Gc_t$  defined in (\ref{eq:NONCVX}) with $T=W$, \ie $\Gc^{\RED} := (\Gc_1,\cdots,\Gc_T)$. The look-ahead horizon of each rolling-window is defined by $\Hc_t =\{t,\cdots, t+W-1\}$. The interval $t$ is called the {\em binding interval} and the rest of  $\Hc_t$  the {\em advisory intervals.}  As time $t$ increases, $\Hc_t$  slides across the entire scheduling period $\Hc$. Let $(g_{it}^{{\mbox{\tiny D}}*},g_{it}^{{\mbox{\tiny C}}*})$ and $(\lambda_{t}^{*})$ be the solution to (\ref{eq:NONCVX}). The rolling-window dispatch, $\Gbf^{\RED}:=\{\Gbf^{{\RED}\mbox{\tiny-D}},\Gbf^{{\RED}\mbox{\tiny-C}}\}$, and rolling-window LMP (R-LMP), $\pibf^{\mbox{\rm\tiny R-LMP}}$, in interval $t$ are given by policy $\Gc^{\RED}$ with
\beq \label{eq:gED}
\begin{array}{c}
g_{it}^{\RED{\mbox{\tiny-D}}}:= g_{it}^{{\mbox{\tiny D}}*},~g_{it}^{\RED{\mbox{\tiny-C}}}:= g_{it}^{{\mbox{\tiny C}}*},~\pi^{\mbox{\rm\tiny R-LMP}}_{t} := \lambda_{t}^{*}, \forall i.
\end{array}
\eeq
This means, in each rolling window, the binding interval dispatch and pricing signals are implemented, and those signals for advisory intervals are not guaranteed to be materialized.

\subsection{LOC as a measure of individual rationality}
The  lost opportunity costs (LOC) payment of individual storage is a measure of the individual rationality (or the dispatch-following incentive), defined by the difference between the payment that would have been received had the storage self-scheduled and the payment received within the market clearing process. Let  $\pibf=(\pi_1,\cdots,\pi_T)$ be the column vector of a realized  uniform price over the entire scheduling horizon $\Hc$. Denote $\gbf^{\mbox{\tiny R-ED}}:=\{\gbf^{\mbox{\tiny R-ED-C}}, \gbf^{\mbox{\tiny R-ED-D}}\}$. We here omit storage index $i$ for simplicity. And the LOC over the scheduling horizon $\Hc$ is given by 
\beq \label{eq:uplift}
\begin{array}{c}
{\rm LOC}(\pibf,\gbf^{\mbox{\tiny R-ED}})=Q(\pibf) - \pibf^\intercal (\gbf^{\mbox{\tiny R-ED-D}}- \gbf^{\mbox{\tiny R-ED-C}})+F(\gbf^{\mbox{\tiny R-ED}};s),
\end{array} 
\eeq
where $Q(\pibf)$ is the maximum profit the storage would have received through the individual profit maximization  (\ref{eq:Q}). Note that Theorem.~\ref{thm:LOC_MIO} indicates ${\rm LOC}(\pibf^{\mbox{\rm\tiny LMP}}, \gbf_i^{\mbox{\rm \tiny D}*},\gbf_i^{ \mbox{\rm \tiny C}*})=0, \forall i$ in the one-shot dispatch. But this nice property of LMP fails to be extended to the rolling-window dispatch \cite{Guo&Chen&Tong:21TPS}. Our previous results \cite{ChenTong22EPSRstorage} show that rolling-window temporal locational marginal pricing (R-TLMP) needs {\em zero} LOC payment to support the individual rationality of all market participants with SoC-independent bids, although it's a discriminative price  . What happens if we have SoC-dependent bids? Can R-TLMP still maintain this nice property? The answer is yes and details are explained in the following section.

\subsection{R-TLMP and SoC-dependent bids with end-state control}
TLMP is a non-uniform marginal cost pricing that measures the marginal contribution of the resource to meeting the demand $d_t$ at the optimal dispatch. Let
$\bar{F}^{(-i)}_t(\Gbf_t^*, \sbf)$ be the total cost in rolling-window $t$, excluding the contribution of storage $i$ in interval $t$  by treating $(g_{it}^{\mbox{\tiny C*}},g_{it}^{\mbox{\tiny D*}})$  as parameters set at the optimal dispatch point, i.e.   
\beq
\begin{array}{lrl}
\bar{F}^{(-i)}_t(\Gbf^*, \sbf)&=&\bar{F}(\Gbf^*, \sbf)-(f^{\mbox{\tiny D}}_{it} (g^{\mbox{\tiny D*}}_{it})- f^{\mbox{\tiny C}}_{it} (g^{\mbox{\tiny C*}}_{it})).
\end{array}
\eeq
With envelope theory, the marginal contribution from storage $i$ in interval $t$ can be computed with the optimal primal and dual solutions of  (\ref{eq:NONCVX}) by
\bea \label{eq:RTLMPESR}
\pi_{it}^{\mbox{\tiny TLMP}}&:=&
 \left\{\begin{array}{ll}
 \frac{\partial}{\partial g^{\mbox{\tiny C}}_{it}}  \bar{F}^{(-i)}_t(\Gbf^*, \sbf) & \mbox{charging}\\
 - \frac{\partial}{\partial g^{\mbox{\tiny D}}_{it}}  \bar{F}^{(-i)}_t(\Gbf^*, \sbf) & \mbox{discharging}\\
 \end{array}\right.\\
&=&
  \left\{\begin{array}{ll}
 \lambda^*_t- \eta_i^{\mbox{\tiny C}}\phi^*_{it}-\Delta_{it}^{\mbox{\tiny C}*} :=\pi_{it}^{\mbox{\tiny C}}, & \mbox{charging}\\
 \lambda^*_t- 1/\eta_i^{\mbox{\tiny D}}\phi^*_{it}+\Delta_{it}^{\mbox{\tiny D}*}:=\pi_{it}^{\mbox{\tiny D}}, & \mbox{discharging}\\
    \end{array}
    \right.
\eea
where $\Delta_{it}^{\mbox{\tiny C}*}$ (and, similarly,  $\Delta_{it}^{\mbox{\tiny D}*}$) is  defined by
\bea
\Delta_{it}^{\mbox{\tiny C}*}&:=&( \bar{\mu}_{i(t+1)}^{\mbox{\tiny C}*}-\underline{\mu}_{i(t+1)}^{\mbox{\tiny C}*})-(\bar{\mu}_{it}^{\mbox{\tiny C}*}-\underline{\mu}_{it}^{\mbox{\tiny C}*}) .
\eea
TLMP prices the inelastic demand with the same way as LMP. Note that TLMP $\pi_{it}^{\mbox{\tiny TLMP}}$ for storage $i$ can be decomposed into the energy price $\pi^{\mbox{\rm\tiny LMP}}_{t}=\lambda^*_t$, the SOC price $\phi^*_{it}$, and ramping prices $(\Delta_{it}^{\mbox{\tiny C}*}, \Delta_{it}^{\mbox{\tiny D}*})$. When there is no binding ramping and SOC constraints, TLMP reduces to LMP. The above formulation of TLMP is in parallel with that in \cite{ChenTong22EPSRstorage} since the multi-interval dispatch is convex with the EDCR condition in each rolling window.

Corresponding to the rolling-window pricing  rules in (\ref{eq:gED}), we have R-TLMP $\pibf^{\mbox{\rm\tiny R-TLMP}}_{i}:=(\pibf_i^{\mbox{\tiny D}},\pibf_i^{\mbox{\tiny C}} )$ in interval $t$ given by policy $\Gc^{\RED}$ with
\beq \label{eq:RTLMP}
\begin{array}{c}
\pi^{\mbox{\rm\tiny R-TLMP}}_{it} := \pi_{it}^{\mbox{\tiny TLMP}}.
\end{array}
\eeq

The following theorem establishes that, under R-TLMP, the LOC for every storage is zero with the end-state SoC control $e_{\mbox{\rm \tiny T+1}} \in \Ec_{\gamma}$, which requires the end-state SoC falls into certain SoC partition $ \Ec_{\gamma}$  for a given $\gamma$. And it is locally optimal that every price-taker bids truthfully under R-TLMP in the rolling-window dispatch.

\begin{theorem}[Individual rationality and truthful-bidding incentive under R-TLMP] \label{thm:LOC_RW}   For storage $i$, let $\gbf^{\mbox{\tiny R-ED}}_i$ be the rolling-window economic dispatch computed from (\ref{eq:gED}) and  $\pibf^{\mbox{\rm\tiny R-TLMP}}_{i}$ be its R-TLMP from (\ref{eq:RTLMP}). With end-state SoC control $e_{\mbox{\rm \tiny T+1}}\in \Ec_{\gamma}$, we have
 \beq \label{eq:StorageLOC=0}
\mbox{\rm LOC}(\pibf^{\mbox{\rm\tiny R-TLMP}}_{i},\gbf^{\mbox{\tiny R-ED}}_i)=0, \forall i,
\eeq
when the EDCR condition and Assumption~\ref{assume:single}  are satisfied. And it is optimal for a price-taking storage to bid truthfully with its marginal costs of charging and discharging. 
\end{theorem}
{\em Proof:}   See Appendix for the proof.

From Theorem~\ref{thm:Eq}, we know that with the end-sate SoC control, \ie $e_{\mbox{\rm \tiny T+1}} \in \Ec_{\gamma}$, the cost of storage over multi-interval is a linear function rather than piecewise linear. So, in each rolling-window solving economic dispatch (\ref{eq:NONCVX}), the piecewise linear cost function $F(\gbf^{\mbox{\tiny C}}, \gbf^{\mbox{\tiny D}};s)$ is replaced by the linear function $F(\gbf^{\mbox{\tiny C}}, \gbf^{\mbox{\tiny D}};s, \gamma)$. So does the individual profit maximization in (\ref{eq:Q}). Therefore, with the end-state control, Theorem~\ref{thm:LOC_RW} can be proved based on the convexity credit to the EDCR condition and the linear objective functions. 

The truthful-bidding incentive follows the individual rationality property. Known that $\mbox{\rm LOC}$ is zero and the rolling window dispatch signal is an optimal solution for (\ref{eq:Q}), truthful-bidding storage will receive the rolling window dispatch signal that is optimal for individual profit maximization.

\section{A Numerical Example} \label{sec:NE}
The individual rationality of one-shot LMP stated in Theorem~\ref{thm:LOC_MIO}  guarantees that the individual optimal dispatch from  (\ref{eq:Q}) is optimal to the centrally one-shot dispatch (\ref{eq:NONCVX}) if they use/reach the same one-shot LMP.  So, we establish the following numerical example with exogenous LMP to analyze the mileage, profit, and profit margin of storage with different bid-in costs. This numerical example is equivalent to the simulation of a central market clearing model with storage and other market participants reaching the same LMP as the adopted exogenous LMP scenarios.

\begin{figure}[h]
\center
\begin{psfrags}
\scalefig{0.24}\epsfbox{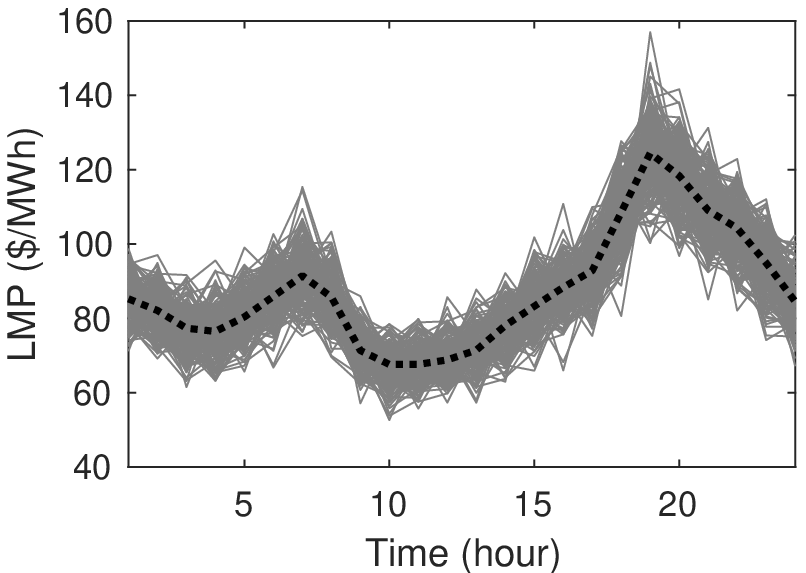}\scalefig{0.24}\epsfbox{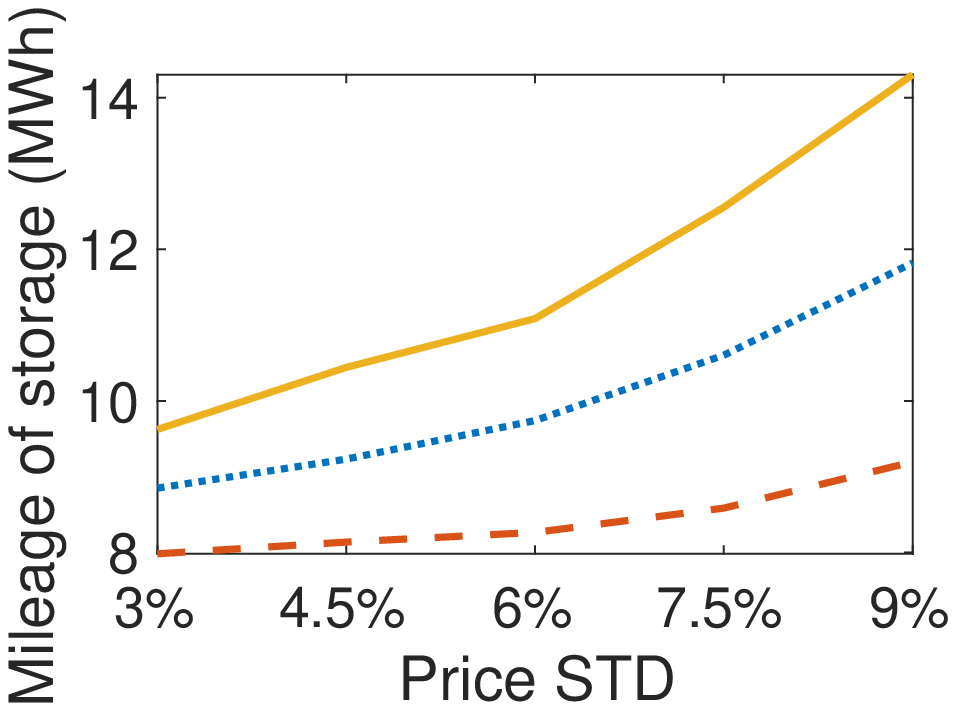}
\scalefig{0.24}\epsfbox{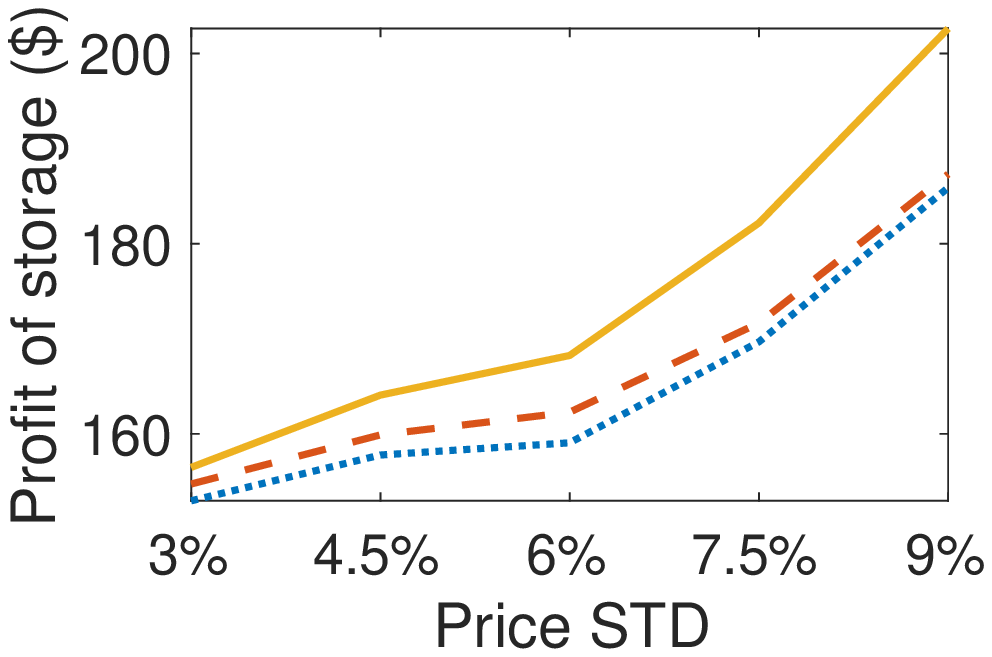}\scalefig{0.24}\epsfbox{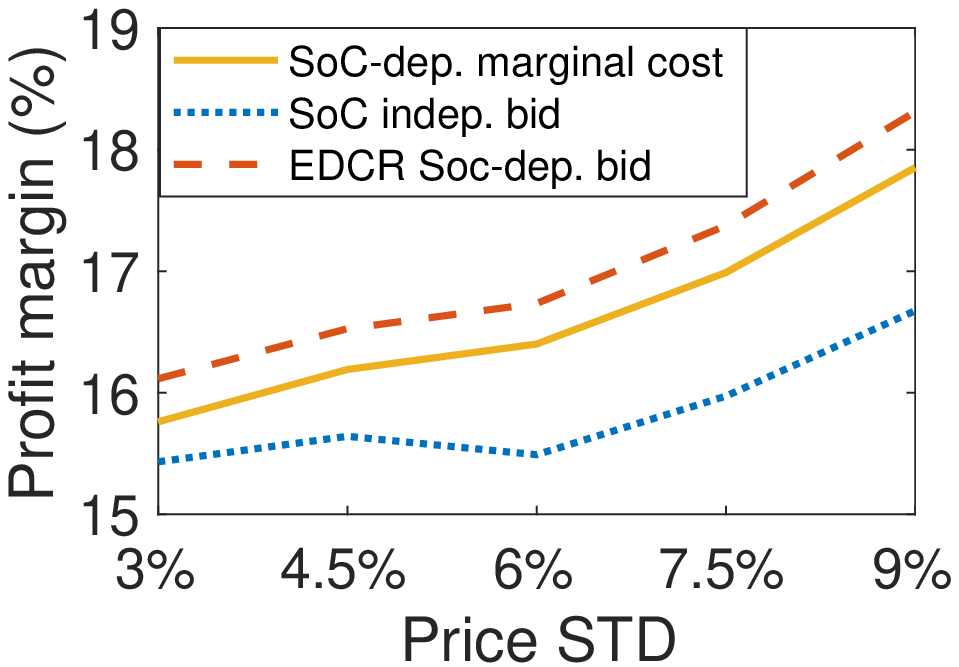}
\end{psfrags}
\vspace{-0.5em}\caption{\scriptsize Top left: exogenous LMP scenarios (dash line represents mean value). Top right: Mileage of storage vs. price STD. Bottom left: profit of storage vs. price STD. Bottom right: Profit margin vs. price STD. }
\label{fig:ProfitMargin}
\end{figure}

Consider ideal storage in the 2-interval dispatch with initial SoC $s_1=17.5$ {\rm MWh}. For the true SoC-dependent marginal cost, we have $K=2$, $\cbf^{\mbox{\rm \tiny C}}=(40.3, 9.3)$ {\rm \$/MWh}, $\cbf^{\mbox{\rm \tiny D}}=(106.7, 50.7)$ {\rm \$/MWh}, and $\Ebf=(9, 20, 25)$MWh. And the optimal EDSR SoC-dependent bid computed by the method from \cite{ChenTong2022convexifying} has  $\tilde{\cbf}^{\mbox{\rm \tiny C}}=\cbf^{\mbox{\rm \tiny C}}$ {\rm \$/MWh}, $\tilde{\cbf}^{\mbox{\rm \tiny D}}=(106.7, 75.7)$ {\rm \$/MWh}. Let the capacity limit of charge and discharge power be $ \bar{g}^{\mbox{\rm \tiny C}}= \bar{g}^{\mbox{\rm \tiny D}}=5$MW, and ramping limits are ignored. We compare the mileage, profit, and profit margin of storage with different bids including the true SoC-dependent marginal cost, the SoC-independent bid, and the EDCR SoC-dependent bid, into the electricity market. And the market is cleared by one-shot dispatch $\Gc$ and LMP $\pibf^{\mbox{\tiny LMP}}$.  Note that the optimal SoC-independent bid is also computed by the EDCR approximation method from \cite{ChenTong2022convexifying}, and the result is 88.94\$/MWh for discharge power and 30.47\$/MWh for charge power. We solve the optimal profit of storage (\ref{eq:Q}) under 200 exogenous LMP scenarios, which are generated from a  CAISO baseline scenario \cite{CAISOLMP} with the normalized standard deviation (STD) varying from 3\% to 9\%. The pricing scenarios are shown in Fig.~\ref{fig:ProfitMargin}  (top left).

Simulation results related to the average mileage, average profit, and average profit margin of storage are respectively shown in the top right, bottom left, and bottom right of  Fig.~\ref{fig:ProfitMargin}. Profit of storage is computed by the objective value of (\ref{eq:Q}), where we have revenue minus storage cost. And profit margin is computed by profit divided by the revenue of storage. We  can observe that the storage is less frequently used with the EDCR SoC-dependent bid. And the profit of storage with EDCR bid is less than the true SoC-dependent marginal cost. However, the storage has the highest average profit margin with the EDCR SoC-dependent bid. As for the SoC-independent bid, which is adopted in the current electricity market, it fails to give an accurate approximation of the true SoC-dependent marginal cost in this example, and the storage doesn't receive an economic enough operation with the SoC-independent bid.

\section{Conclusion}\label{sec:conclusion}
The nonconvexity of the multi-interval market clearing problem resulting from the SoC-dependent bid is analyzed in this paper, and we propose a sufficient convexification condition on the bidding format, the equal decremental-cost ratio (EDCR) condition, to transform the market clearing problem into a standard convex piecewise linear program. Such EDCR SoC-dependent bids convexify the economic dispatch, reducing the integer programming problem to a linear program, reducing the computation burden of including large-scale storage deployment, and requiring minimal changes in the market clearing and pricing engine. 

With the convex multi-interval dispatch preserved by the EDCR condition, we further establish the individual rationality for storage under one-shot LMP and rolling-window TLMP. Therefore, we can price storage operation with the dispatch-following incentive support, and eliminates the need for out-of-the-market uplifts. 

Many relevant issues are outside the scope of this work, requiring further investigation. One is deriving optimal EDCR bids based on SoC-dependent costs. To this end, the preliminary work in \cite{ChenTong2022convexifying} offers a possible solution. The other is the generalization of the results presented here to time-varying EDCR bids. Finally, more extensive simulations and empirical analysis are needed to evaluate the costs and benefits of SoC-dependent bids. 


 \section*{Acknowledgement}
The authors are grateful for discussions with Tongxin Zheng and Benjamin Hobbs.  
{
\bibliographystyle{IEEEtran}
\bibliography{BIB}
}

\section*{Appendix}
\subsection{Proof of Theorem~\ref{thm:LOC_MIO} }
The following proof comes directly from the strong duality between the primal and dual problem of the one-shot dispatch (\ref{eq:NONCVX}) because the one-shot dispatch (\ref{eq:NONCVX}) is convex with the EDCR condition satisfied. 

The individual profit maximization problem (\ref{eq:Q}) with complete dual variables is shown below
\beq \label{eq:Q_complete}
\begin{array}{lcl}
&Q(\pibf)= \underset{\{\pbf^{\mbox{\tiny D}},\pbf^{\mbox{\tiny C}},\ebf\}}{\rm maximize} &
 \pibf^{\T}\pbf^{\mbox{\tiny D}}-\pibf^{\T}\pbf^{\mbox{\tiny C}}-F(\pbf^{\mbox{\tiny C}}, \pbf^{\mbox{\tiny D}};s)\\
& {\rm subject~to}&\\
&\psibf: & \eta^{\mbox{\tiny C}}\pbf^{\mbox{\tiny C}}-\pbf^{\mbox{\tiny D}}/\eta^{\mbox{\tiny D}}=\Abf\qbf, \\
& (\underline{\xibf}^{\mbox{\tiny D}},\bar{\xibf}^{\mbox{\tiny D}}):  & -\underline{\rbf}^{\mbox{\tiny D}}\le \Bbf \pbf^{\mbox{\tiny D}} \le \bar{\rbf}^{\mbox{\tiny D}}, \\
& (\underline{\xibf}^{\mbox{\tiny C}},\bar{\xibf}^{\mbox{\tiny C}}):  & -\underline{\rbf}^{\mbox{\tiny C}}\le \Bbf \pbf^{\mbox{\tiny C}} \le \bar{\rbf}^{\mbox{\tiny C}}, \\
 &(\underline{\zetabf}^{\mbox{\tiny D}},\bar{\zetabf}^{\mbox{\tiny D}}):&{\bf 0}\leq \pbf^{\mbox{\tiny D}} \leq\bar{\gbf}^{\mbox{\tiny D}},\\
 &(\underline{\zetabf}^{\mbox{\tiny C}},\bar{\zetabf}^{\mbox{\tiny C}}):&{\bf 0}\leq \pbf^{\mbox{\tiny C}} \leq \bar{\gbf}^{\mbox{\tiny C}},\\
&& \underline{\ebf}\le \qbf \le \bar{\ebf}, 
\end{array} \hfill
\eeq

And the individual optimal dispatch computed from (\ref{eq:Q}) satisfies KKT conditions\footnote{Subgradients are used for nondifferentiable points.}:
\beq \label{eq:StorageLOCKKT}
\begin{array}{l}
\nabla_{\tiny \pbf^{\mbox{\tiny D}}} F(\cdot)-\pibf+1/\eta^{\mbox{\tiny D}}\psibf^* -\chibf^{\mbox{\tiny D}*}+ \Delta\zetabf^{{\mbox{\tiny D}}*} ={\bf 0},\\
-\nabla_{\tiny \pbf^{\mbox{\tiny C}}} F(\cdot)+\pibf-\eta^{\mbox{\tiny C}}\psibf^*-\chibf^{\mbox{\tiny C}*}+ \Delta\zetabf^{{\mbox{\tiny C}}*} ={\bf 0},
\end{array}
\eeq
where we have $\chi_{t}^{\mbox{\tiny D}*}:=\Delta \xi_{t+1}^{{\mbox{\tiny D}}*}-\Delta \xi_{t}^{{\mbox{\tiny D}}*}$, $\Delta \xi_{t}^{\mbox{\tiny D}*}:=\bar{\xi}_{t}^{\mbox{\tiny D}*}-\underline{\xi}_{t}^{\mbox{\tiny D}*}$, and $\Delta\zetabf^{\mbox{\tiny D}*}=\bar{\zetabf}^{\mbox{\tiny D}*} - \underline{\zetabf}^{\mbox{\tiny D}*}$ (and, similarly, $\Delta\zetabf^{\mbox{\tiny C}*}$, $\chi_{t}^{\mbox{\tiny C}*}$).

The compact form of the one-shot dispatch problem (\ref{eq:NONCVX}) is
\beq \label{eq:NONCVX_vector}
\begin{array}{lrl}
\Gc^{\tiny\rm ED}: &\underset{\{(g_{it}^{\mbox{\tiny C}}, g_{it}^{\mbox{\tiny D}}, e_{it})\}}{\rm minimize} &  \bar{F}(\Gbf, \sbf)
\\& s.t.&  \forall i\in \{1,...,N\}\\
&&\underline{\ebf}_i \le \ebf_{i}\le \bar{\ebf}_i,\ebf_{i}=\sbf_i,\\
&\phibf_{i}:& \gbf^{\mbox{\tiny C}}_{i}\eta^{\mbox{\tiny C}}-\gbf^{\mbox{\tiny D}}_{it}/\eta^{\mbox{\tiny D}}=\Abf\ebf,\\
&(\underline{\rhobf}^{\mbox{\tiny C}}_i,\bar{\rhobf}^{\mbox{\tiny C}}_i):& {\bf 0} \le \gbf^{\mbox{\tiny C}}_{i}\le \bar{\gbf}^{\mbox{\tiny C}}_i,\\
&(\underline{\rhobf}^{\mbox{\tiny D}}_i,\bar{\rhobf}^{\mbox{\tiny D}}_i):& {\bf 0} \le \gbf^{\mbox{\tiny D}}_{i}\le \bar{\gbf}^{\mbox{\tiny D}}_i,\\
&(\underline{\mubf}^{\mbox{\tiny C}}_i,\bar{\mubf}^{\mbox{\tiny C}}_i):& -\underline{\rbf}^{\mbox{\tiny C}}_i \le \Bbf\gbf^{\mbox{\tiny C}}_{i}\le \bar{\rbf}^{\mbox{\tiny C}}_i,\\
&(\underline{\mubf}^{\mbox{\tiny D}}_i,\bar{\mubf}^{\mbox{\tiny D}}_i):& -\underline{\rbf}^{\mbox{\tiny D}}_i  \le \Bbf\gbf^{\mbox{\tiny D}}_{i}\le  \bar{\rbf}^{\mbox{\tiny D}}_i,\\
&\lambdabf:& \sum_{i=1}^{N}(\gbf^{\mbox{\tiny D}}_{i}-\gbf^{\mbox{\tiny C}}_{i})=\hat{\dbf}.
\end{array}
\eeq

And the optimal solution satisfies KKT conditions:
\beq \label{eq:DispatchKKTSO}
\begin{array}{l}
\nabla_{\tiny \gbf^{\mbox{\tiny D}}} F(\cdot)- \lambdabf^*+1/\eta^{\mbox{\tiny D}} \phibf_{i}^* -\Deltabf_{i}^{\mbox{\tiny D}*}+ \Delta\rhobf^{{\mbox{\tiny D}}*}_{i}= 0,\forall i\\
-\nabla_{\tiny \gbf^{\mbox{\tiny C}}} F(\cdot)+ \lambdabf^* -\eta^{\mbox{\tiny C}} \phibf_{i}^*-\Deltabf_{i}^{\mbox{\tiny C}*}+ \Delta\rhobf^{{\mbox{\tiny C}}*}_{i}= 0,\forall i,
\end{array}
\eeq
where $\Delta_{it}^{\mbox{\tiny D}*}:= \Delta \mu_{i(t+1)}^{{\mbox{\tiny D}}*}-\Delta \mu_{it}^{{\mbox{\tiny D}}*}$, $\Delta \mu_{it}^{\mbox{\tiny D}*}:=\bar{\mu}_{it}^{\mbox{\tiny D}*}-\underline{\mu}_{it}^{\mbox{\tiny D}*}$, and $\Delta\rho^{\mbox{\tiny D}*}_{it}:=\bar{\rho}^{\mbox{\tiny D}*}_{it} - \underline{\rho}^{\mbox{\tiny D}*}_{it}$ (and, similarly, $\Delta_{it}^{\mbox{\tiny C}*}, \Delta \mu_{it}^{\mbox{\tiny C}*}$). By definition, we have $\pi^{\mbox{\rm\tiny LMP}}_t=\lambda^*_t$.


The KKT conditions of the individual optimization shown in (\ref{eq:StorageLOCKKT}) for all storage $i$ and KKT conditions of one-shot dispatch  (\ref{eq:DispatchKKTSO}) can be simultaneously satisfied by setting $\pi_t =\pi^{\mbox{\rm\tiny LMP}}_t$, $g^{\mbox{\tiny D}*}_{it}= p^{{\mbox{\tiny D}}*}_{it},\bar{\rho}^{\mbox{\tiny D}*}_{it}=\bar{\zeta}^{\mbox{\tiny D}*}_{it},\underline{\rho}^{\mbox{\tiny C}*}_{it}=\underline{\zeta}^{\mbox{\tiny C}*}_{it},\bar{\xi}_{it}^{\mbox{\tiny D}*}=\bar{\mu}_{it}^{\mbox{\tiny D}*},\underline{\xi}_{it}^{\mbox{\tiny D}*}=\bar{\mu}_{it}^{\mbox{\tiny D}*},  \phi^*_{it}=\psi^*_{it},\forall i,t$ (The same group of equations are applied for variables with superscript $C$). And we have ${\rm LOC}(\pibf^{\mbox{\rm\tiny LMP}}, \gbf_i^{\mbox{\rm \tiny D}*},\gbf_i^{ \mbox{\rm \tiny C}*})=0$ from the LOC definition (\ref{eq:uplift}).  

Next, we prove the truthful-bidding incentive for price takers under one-shot LMP. Let $\thetabf$ be the bidding parameter of the storage in the wholesale market. Under the price taker assumption, each individual storage has profit given by
\beq
\begin{array}{lrl}\label{eq:BidTrue}
\Pi(\thetabf^*)&=&(\pibf^{\mbox{\rm\tiny LMP}})^{\intercal}(\gbf^{\mbox{\rm \tiny D}*}(\thetabf^*)-\gbf^{\mbox{\rm \tiny C}*}(\thetabf^*))\\
&&-F(\gbf^{\mbox{\rm \tiny C}*}(\thetabf^*), \gbf^{\mbox{\rm \tiny D}*}(\thetabf^*);\sbf, \thetabf^*)\\
 &\geq& (\pibf^{\mbox{\rm\tiny LMP}})^{\intercal}(\gbf^{\mbox{\rm \tiny D}*}(\thetabf)-\gbf_i^{\mbox{\rm \tiny C}*}(\thetabf))\\
&&-F(\gbf^{\mbox{\rm \tiny C}*}(\thetabf), \gbf^{\mbox{\rm \tiny D}*}(\thetabf);\sbf, \thetabf),\\
\end{array}
\eeq
where $\thetabf^*$ is the the truthful bidding parameter. The second inequality follows that there will be no LOC under the one-shot LMP. So, the storage has optimal profit when bid truthfully.  \QED

\subsection{Proof of Theorem~\ref{thm:LOC_RW}} 
With the end-sate SoC control, \ie $e_{\mbox{\rm \tiny T+1}} \in \Ec_{\gamma}$, the multi-interval cost of storage is a linear function rather than piecewise linear (From Theorem~\ref{thm:Eq}). So, in each rolling-window solving multi-interval look-ahead dispatch (\ref{eq:NONCVX_vector}), the piecewise linear cost function $F(\gbf^{\mbox{\tiny C}}, \gbf^{\mbox{\tiny D}};s)$ is replaced by the linear function $F(\gbf^{\mbox{\tiny C}}, \gbf^{\mbox{\tiny D}};s, \gamma)$. So does the individual profit maximization in (\ref{eq:Q_complete}).

Also note that KKT conditions for the rolling-window dispatch, \ie (\ref{eq:DispatchKKTSO}), come from the binding interval of each rolling-window optimization rather than a single multi-interval dispatch problem. We here show that KKT conditions of the individual profit maximization (\ref{eq:StorageLOCKKT}) and the rolling-window dispatch can be simultaneously satisfied by setting $\pi_t=\pi^{\mbox{\rm\tiny R-TLMP}}_{it}, g^{\mbox{\tiny R-ED-D}}_{it}= p^{{\mbox{\tiny D}}*}_{it},g^{\mbox{\tiny R-ED-C}}_{it}= p^{{\mbox{\tiny C}}*}_{it},\Delta\rho^{\mbox{\tiny D}*}_{it}=\Delta\zeta^{\mbox{\tiny D}*}_{it},\Delta\rho^{\mbox{\tiny C}*}_{it}=\Delta\zeta^{\mbox{\tiny C}*}_{it},\bar{\xi}_{it}^{\mbox{\tiny D}*}=\underline{\xi}^{\mbox{\tiny D}*}_{it}=\bar{\xi}_{it}^{\mbox{\tiny C}*}=\underline{\xi}^{\mbox{\tiny C}*}_{it}=\psi^*_{it}=0,\forall i,t$. That way, we have $\mbox{\rm LOC}(\pibf^{\mbox{\rm\tiny R-TLMP}}_{i},\gbf^{\mbox{\tiny R-ED}}_i)=0, \forall i$ from the definition of LOC (\ref{eq:uplift}).

Since we have zero LOC under the R-TLMP  for each storage participant, the truthful-bidding incentive can be similarly proved by using $\pibf^{\mbox{\rm\tiny R-TLMP}}_i$ and rolling-window dispatch signals $\gbf^{\mbox{\tiny R-ED}}_i$ in equation (\ref{eq:BidTrue}). \QED


\edoc

\end{document}